\documentclass[11pt,pdftex]{article}

\usepackage{amssymb}
\usepackage{amsmath}
\usepackage{color} 
\usepackage{fullpage}
\ifx\pdftexversion\undefined
\usepackage[dvips]{graphicx}
\else
  \usepackage[pdftex]{graphicx}
\fi

\newtheorem{theorem}{Theorem}
\newtheorem{lemma}[theorem]{Lemma}
\newenvironment{proof}[1][Proof]{\noindent\textbf{#1.} }{\hfill $\Box$\\[2mm]} 

\def\lf{\tiny}

\def\nnll{\refstepcounter{linenumber}\lf\thelinenumber}
\newcounter{linenumber}

\def\A{\ensuremath{\mathcal{A}}}

\def\I{\ensuremath{\mathcal{I}}}
\def\O{\ensuremath{\mathcal{O}}}

\def\L{\ensuremath{\mathcal{L}}}

\def\T{\ensuremath{\mathcal{T}}}

\def\val{\textit{val}}
\def\code{\textit{code}}

\newcommand{\correct}{\mathit{correct}}

\newcommand{\true}{\mathit{true}}
\newcommand{\false}{\mathit{false}}

\newcommand{\remove}[1]{}

\newcommand{\id}[1]{\mbox{\textit{#1}}}
\newcommand{\HS}{\textit{HS}}
\newcommand{\HSS}{h}

\newcommand{\ignore}[1]{}

\begin{document}

\bibliographystyle{abbrv}

\title{Relating $\L$-Resilience and Wait-Freedom \\via Hitting Sets}

\author{Eli Gafni\\
\\
\normalsize Computer Science Department, UCLA\\
\and
Petr Kuznetsov\thanks{Contact author: pkuznets@acm.org, TU Berlin, Sekr. TEL 16,
  Ernst-Reuter-Platz 7, 10587 Berlin, Fax: +49 391 534 783 47}\\
\\
\normalsize Deutsche Telekom Laboratories/TU Berlin\\
}

\date{}
\maketitle

\begin{abstract}
The condition of $t$-resilience 
stipulates that an $n$-process program is only obliged to make progress 
when at least $n-t$ processes are correct. 
Put another way, the \emph{live sets},
the collection of process sets such that progress is required if
all the processes in one of these sets are correct, are all sets with at least $n-t$ processes. 

We show that the ability of arbitrary collection of live sets $\L$ to
solve distributed tasks 
is tightly related to the \emph{minimum hitting set} 
of $\L$, a minimum cardinality subset of processes that has a non-empty 
intersection with every live set.
Thus, finding the computing power of $\L$ is $NP$-complete.

For the special case of \emph{colorless} tasks that allow
participating processes to adopt input or output values of each other,
we use a simple simulation to show  that a task can be solved
\emph{$\L$-resiliently}  if
and only if it can be solved $(h-1)$-resiliently, where $h$ is the
size of the minimum hitting set of $\L$.

For general tasks, 
 we characterize $\L$-resilient solvability 
of tasks with respect to a limited notion of 
\emph{weak} solvability:  in every execution where all processes in some set in $\L$ are correct, 
outputs must be produced for every process in some (possibly
different) participating set in $\L$.
Given a task $T$, we construct another task $T_{\L}$ such that $T$
is solvable weakly $\L$-resiliently if and only if $T_{\L}$
is solvable weakly wait-free.
\end{abstract}




\section{Introduction}

One of the most intriguing questions in distributed computing 
is how to distinguish solvable from the unsolvable.
Consider, for instance, the question of \emph{wait-free} solvability 
of distributed tasks. 
Wait-freedom does not impose any restrictions on the 
scope of considered executions, i.e., 
a wait-free solution to a task requires every correct processes to output in every execution. 
However, most interesting distributed tasks cannot be solved
in a wait-free manner~\cite{FLP85,LA87}. 
Therefore, much research is devoted to understanding how 
the power of solving a task increases as the scope of considered
executions decreases.
For example, \emph{$t$-resilience} 
considers only executions where at least $n-t$ processes are correct 
(take infinitely many steps), where $n$ is the number of processes in the system.
This provides for solving a larger set of tasks than wait-freedom,
since in executions in which less than $n-t$ processes are correct, 
no correct process is required to output.

What tasks are solvable $t$-resiliently? It is known that this question is undecidable even 
with respect to wait-free solvability, let alone $t$-resilient~\cite{GK99-undecidable,HR97}. 
But is the question about
$t$-resilient solvability in any sense different than the question about wait-free solvability?
If we agree that we ``understand'' wait-freedom~\cite{HS99}, do we understand $t$-resilience to a lesser degree? 
The answer should be a resounding no if,  in the sense of solving tasks, the models
can be reduced to each other. That is, if for every task $T$ we
can find a task $T_{t}$ which is solvable wait-free if and only 
if $T$ is solvable $t$-resiliently.
Indeed, \cite{BG93b,BGLR01,Gaf09-EBG} established that $t$-resilience can be reduced to wait-freedom.
Consequently, the two models are unified with respect to task solvability. 

In this paper, we consider a generalization of $t$-resilience, 
called \emph{$\L$-resilience}.
Here $\L$ stands for a collection of subsets of processes. A set in $\L$ is referred to as 
a \emph{live set}.
In the model of $\L$-resilience, a correct process is only obliged to
produce outputs if all the processes in some live set are correct. 
Therefore, the notion of $\L$-resilience represents a restricted class of
\emph{adversaries} introduced by Delporte et
al.~\cite{delporte09adversary}, 
described as collections of exact correct sets.
$\L$-resilience describes adversaries that are closed under the
superset operation: 
if a correct set is in an adversary, then every superset of it is also
in the adversary.

We show that the key to understanding $\L$-resilience is the notion of
a \emph{minimum hitting set} of $\L$ (called simply hitting set in the
rest of the paper).
Given a set system $(\Pi,\L)$ where $\Pi$ is a set of processes and $\L$ is a set of subsets
of $\Pi$,  $H$ is a hitting set of $(\Pi,\L)$ if it is a minimum cardinality subset of $\Pi$ 
that meets every set in $\L$. 
Intuitively, in every $\L$-resilient execution, i.e., in every execution in which at least one 
set in $\L$ is correct, not all processes in a hitting set of $\L$ can
fail.
Thus, under $\L$-resilience, we can solve  
the $k$-set agreement task among the processes in $\Pi$ 
where $k$ is the hitting set size of $(\Pi,\L)$. 
In $k$-set agreement, the processes start with 
private inputs and the set of outputs is a subset of inputs of size at most $k$.
Indeed, fix a hitting set $H$ of $(\Pi,\L)$ of size $k$. Every process in $H$ simply
posts its input value in the shared memory, and every other process returns the first value
it witnesses to be posted by a process in $H$.
Moreover,  using a simple simulation based on~\cite{BG93b,BGLR01}, 
we derive that $\L$ does not allow solving $(k-1)$-set agreement
or any other \emph{colorless} task that cannot be solved
$(k-1)$-resiliently.
Thus, we can decompose superset-closed adversaries into
equivalence classes, one for each hitting set size,
where each class agrees on the set of colorless tasks it allows 
for solving.

Informally, colorless tasks 
allow a process to adopt an input or output value of any other participating process. 
This restriction gives rise to simulation techniques in which dedicated simulators
independently  ``install'' inputs for other, possibly
non-participating processes, and then take steps on their behalf 
so that the resulting outputs are still correct and can be adopted by
any participant~\cite{BG93b,BGLR01}. 
The ability to do this is a strong simplifying assumption when solvability is analyzed.


For the case of \emph{general} tasks, where inputs cannot be installed
independently,
the situation is less trivial. 
We address general tasks by considering a restricted notion of
\emph{weak solvability}, that requires every execution 
where all the processes in some set in $\L$ are correct to produce 
outputs for every process in some (possibly different) participating set in $\L$.
Note that for colorless tasks, weak solvability is equivalent to
regular solvability that requires every correct process to output.

We relate between wait-free solvability and $\L$-resilient
solvability. 
Given a task $T$ and a collection of live sets $\L$, we define a task $T_{\L}$
such that $T$ is weakly solvable $\L$-resiliently if and only if $T_{\L}$ is weakly solvable wait-free.
Therefore, we \emph{characterize}  $\L$-resilient weak solvability, 
as wait-free solvability has already been characterized in~\cite{HS99}.
Not surprisingly, the notion of a hitting set is crucial in determining $T_{\L}$. 

The simulations that relate $T$ and $T_{\L}$ are interesting in their
own right.  
We describe an agreement
protocol, called \emph{Resolver Agreement Protocol} (or RAP),
by which an agreement is immediately achieved if all processes propose the same value, and otherwise 
it is achieved if eventually a single correct process considers itself a
dedicated \emph{resolver}.
This agreement protocol allows for a novel execution model of wait-free
read-write protocols.
The model guarantees that   an arbitrary number of simulators starting with $j$
distinct initial views should \emph{appear} as $j$ independent
simulators and thus a $(j-1)$-resilient execution can be simulated. 

\remove{
A related, independent, paper to appear in PODC 2010~\cite{HR10} analyzes the same
restricted class of adversaries that we do. It derives a characterization of colorless tasks,
as we do.
In fact, deriving our results just with respect to colorless tasks can
be done just with using run-of-the-mill $BG$ simulation~\cite{BG93b}
(Theorem~\ref{th:colorless}), and 
our main technical difficulty involved characterizing regular tasks
with respect to the weak solvability.
The technique of~\cite{HR10} is based on results of modern
combinatorial topology, and contrasts its simplicity to the 
involved arguments of~\cite{delporte09adversary}. 
This looks unfair, as \cite{HR10}~solves a restricted problem.
In contrast, we 
refute the touted need of using topology to obtain simplicity.
Understanding our paper only requires proficiency in bread and butter
of distributed algorithms. 
}

The rest of the paper is organized as follows.
Section~\ref{sec:model} briefly describes our system model. 
Section~\ref{sec:colorless}  presents a simple categorization of colorless tasks.
Section~\ref{sec:wft} formally defines the wait-free counterpart $T_{\L}$ 
to every task $T$.
Section~\ref{sec:rap} describes RAP, the technical core of our main result. 
Sections~\ref{sec:wfl} and~\ref{sec:lwf} present two directions of our equivalence result: 
from wait-freedom to $\L$-resilience and back.
Section~\ref{sec:related} overviews the related work, and Section~\ref{sec:discussion} 
	concludes the paper by discussing implications of our results and open questions.
Most proofs are delegated to the technical report~\cite{GK10-arxiv}.


\section{Model}
\label{sec:model}

We adopt the conventional shared memory model~\cite{Her91}, 
and only describe necessary details. 

\noindent
\textbf{Processes and objects.}
We consider a distributed system composed of a set $\Pi$
of $n$ processes $\{p_1,\ldots,p_{n}\}$ ($n\geq 2$).
Processes communicate by applying atomic operations on a collection of 
	\emph{shared objects}.
In this paper, we assume that the shared objects are registers that
	export only atomic read-write operations.
The shared memory can be accessed using 
atomic snapshot operations~\cite{AADGMS93}.
An \emph{execution} is a pair $(I,\sigma)$ where $I$ 
is an initial state and $\sigma$ is a sequence of process ids.
A process that takes at least one step in an execution is called \emph{participating}.
A process that takes infinitely many steps in an execution 
is said to be \emph{correct}, otherwise, the process is~\emph{faulty}.

\noindent
\textbf{Distributed tasks.}
A \emph{task} is defined through a set $\I$ of input $n$-vectors 
	(one input value for each process, where the value is $\bot$ for 
	a non-participating process), 
	a set $\O$ of output $n$-vectors 
	(one output value for each process, $\bot$ for
        non-terminated processes)
	and a total relation $\Delta$ that associates each input vector with 
	a set of possible output vectors.
A protocol \emph{wait-free} solves a task $T$ 
if in every execution, every correct process eventually outputs,
 and all outputs respect the specification of $T$.


%

\noindent
\textbf{Live sets.}
The \emph{correct set} of an execution $e$, denoted $\correct(e)$ 
is the set of processes that appear infinitely often in $e$.
For a given collection of live sets $\L$, we say that 
an execution $e$ is \emph{$\L$-resilient} if for some $L\in\L$,
$L\subseteq\correct(e)$.
We consider protocols which allow each process to produce output values for
every other \emph{participating} process in the system by posting the
values 
in the shared memory. 
We say that a process \emph{terminates} when its output value is
posted (possibly by a different process).    

\noindent
\textbf{Hitting sets.}
Given a set system $(\Pi,\L)$ where $\L$ is a set of subsets
of $\Pi$, a set $H\subseteq \Pi$ is a hitting set of $(\Pi,\L)$ if it is a minimum cardinality subset of $\Pi$ 
that meets every set in $\L$. 
We denote the set of hitting sets of $(\Pi,\L)$ by $\HS(\Pi,\L)$, 
and the size of a hitting set of $(\Pi,\L)$ by $\HSS(\Pi,\L)$.
%
By $(\Pi',\L)$, $\Pi' \subseteq \Pi$ we denote the set system 
that consists of 
the elements $S\in\L$, such that $S \subseteq  \Pi'$.
\ignore{
Note that, for each $H\in\HS(\Pi',\L)$,  when $\Pi'$ is the set of participating processes, 
if none of the processes in $H$ are correct, then no process is obliged to output. 
Thus, under $\L$-resilience, we can solve  $|H|$-set agreement for 
participating set $\Pi'$.
This observation provides the key to our main result --- determining a
wait-free counter part for every $\L$-resiliently solvable task.
}
\\
\noindent
\textbf{The BG-simulation technique.}
In a \emph{colorless} task (also called \emph{convergence} tasks~\cite{BGLR01})  
processes are free to use each others' input
and output values, so the task can be defined in terms of 
input and output \emph{sets} instead of vectors.

BG-simulation is a technique by 
	which $k+1$ processes $q_1$, $\ldots$, $q_{k+1}$, called \emph{simulators}, can
	wait-free simulate a $k$-resilient execution of any asynchronous $n$-process 
	protocol~\cite{BG93b,BGLR01} solving a colorless task.
The simulation guarantees that each simulated step of every process $p_j$ is either
	eventually agreed on by all simulators, or the step is
        \emph{blocked} forever and one less simulator participates further 
	in the simulation. 
Thus,  as long there is a live simulator, at least $n-k$ simulated processes 
	accept infinitely many simulated steps.
The technique has been later extended to tasks beyond colorless~\cite{Gaf09-EBG}.

\noindent
\textbf{Weak $\L$-resilience.}
An execution is $\L$-resilient if some set in $\L$ contains only correct processes.
We say that a protocol solves a task $T$ \emph{weakly $\L$-resiliently} 
if in every $\L$-resilient execution, every process in \emph{some} participating set $L\in\L$
eventually terminates, and all posted outputs respect the specification of $T$.
In the wait-free case, when $\L$ consists of all $n$ singletons, 
weak $\L$-resilient solvability 
stipulates that at least one participating process must be given an output value
in \emph{every} execution. 

Weak solvability is sufficient to (strongly) solve every colorless task. 
For general tasks, however, weak solvability does not automatically implies strong solvability, 
since it only allows processes to adopt the output value of any terminated process,
and does not impose any conditions on the inputs.

\section{Colorless tasks}
\label{sec:colorless}
%

First recall the formal definition of a colorless task. 
Let $\val(U)$ denote the set of non-$\bot$ values in a vector $U$.
In a colorless task, for all input vectors $I$ and $I'$
and all output vectors $O$ and $O'$, 
such that $(I,O)\in\Delta$, $\val(I')\subseteq\val(I)$,
$\val(O')\subseteq\val(O)$, 
we have $(I',O)\in\Delta$ and $(I,O')\in\Delta$.

\begin{theorem}
\label{th:colorless}
A colorless task $T$ is weakly $\L$-resiliently solvable if and only if $T$
is $(\HSS(\Pi,\L)-1)$-resiliently solvable.
\end{theorem}
\begin{proof}
Let a colorless task $T$ be $(h-1)$-resiliently solvable, where $h=\HSS(\Pi,\L)$, and let 
$A$ be the corresponding algorithm.
Let $H=q_1,\ldots,q_h$ be a hitting set of $(\Pi,\L)$.
Since $H$ is a hitting set of $\L$, in every $\L$-resilient execution,
at least one simulator must be correct.  
Running BG-simulation~\cite{BG93b,BGLR01} of $A$ on these $h$ simulators, 
where each simulator tries to use its input value of $T$ as an input value 
of every simulated process, results in an $h$-resilient 
simulated execution of $A$.
By our assumption, every correct process must decide in this execution.

For the other direction, suppose, by contradiction that $\L$ 
solves a task $T$ that is not possible to solve $(h-1)$-resiliently.
Let $A_{\L}$ be the corresponding protocol.

Consider any $(h-1)$-resilient execution $e$ of $A_{\L}$, and observe that   
$e$ involves infinitely many steps of a set in $\L$. Indeed,
otherwise, there is a hitting set that does not contain at least
$n-h+1$ processes (namely, the processes that appear infinitely
often in $e$), and thus the hitting set size of $\L$ is at most $h-1$. 

Thus, every $(h-1)$-resilient execution is also $\L$-resilient, which implies an $(h-1)$-resilient solution to $T$ --- a contradiction.        
\end{proof}
Theorem~\ref{th:colorless} implies that $\L$-resilient adversaries can 
be categorized into $n$ equivalence classes, class $h$ corresponding to hitting sets of 
size $h$. Note that two adversaries that belong to the same 
class $h$ agree on the set of colorless tasks they are able to solve,
and the set includes $h$-set agreement.

\section{Relating $\L$-resilience and wait-freedom: definitions}
\label{sec:wft}
%
Consider a set system $(\Pi,\L)$ and a task $T=(\I,\O,\Delta)$, where 
$\I$ is a set of input vectors, 
$\O$ is a set of output vectors, and $\Delta$ is a 
total binary relation between them.    
In this section, we define the ``wait-free'' task $T_{\L}=(\I',\O',\Delta')$ that
characterizes $\L$-resilient solvability of $T$.
%
%
The task $T_{\L}$ is also defined for $n$ processes. 
We call the processes solving $T_{\L}$ \emph{simulators} and 
denote them by $s_1,\ldots, s_n$. 

Let $X$ and $X'$ be two $n$-vectors, and $Z_1,\ldots,Z_n$ be subsets of $\Pi$. 
We say that $X'$ is an \emph{image of $X$ with respect to $Z_1,\ldots,Z_n$}
	if $\forall i$, such that $X'[i]\neq\bot$, we have $X'[i]=\{(j,X[j])\}_{j\in Z_i}$.   

Now $T_{\L}=(\I',\O',\Delta')$ guarantees that 
for all $(I',O')\in\Delta'$, there exist $(I,O)\in\Delta$ such that:

\begin{enumerate}
\item[(1)]$\exists S_1,\ldots, S_n\subseteq \Pi$, each containing a set in $\L$: 

	\begin{enumerate} 

	\item[(1a)] $I'$ is an image of $I$ with respect to $S_1,\ldots,S_n$.

	\item[(1b)] 
	$|\{I'[i]\}_i-\{\bot\}|=m$ $\Rightarrow$ $\HSS(\cup_{i,I'[i]\neq\bot}S_i,\L) \geq m$.

	\end{enumerate}

In other words, every process participating in $T_{\L}$ obtains, as an input, 
a set of inputs of $T$ for some live set, 
and all these inputs are consistent with some input vector $I$ of $T$.

Also, if the number of distinct non-$\bot$ inputs to $T_{\L}$ is $m$, then the hitting set 
size of the set of processes that are given inputs of $T$ is at least $m$.

\item[(2)] $\exists U_1,\ldots,U_n$, each containing a set in $\L$: 
	$O'$ is an image of $O$ with respect to 
	$U_1,\ldots,U_n$. 

In other words, the outputs of $T_{\L}$ produced for input vector $I$ 
should be consistent with $O\in\O$ such that $(I,O)\in\Delta$.	

\end{enumerate}	

Intuitively, every group of simulators that share the same input value
will act as a single process. According to the assumptions on the inputs to $T_{\L}$,
the existence of $m$ distinct inputs implies a hitting set of size at least $m$.
The asynchrony among the $m$ groups will be manifested
as at most $m-1$ failures. The failures of at most $m-1$ processes cannot
prevent \emph{all} live sets from terminating, as otherwise the hitting set in (1b)
is of size at most $m-1$. 

\section{Resolver Agreement Protocol}
\label{sec:rap}

We describe the principal building block of our constructions: 
the resolver agreement protocol (RAP).
RAP is similar to consensus, though it is neither
always safe nor always live. 
To improve liveness, some process may at some point become a
\emph{resolver}, i.e., take the responsibility of making sure that
every correct process outputs.  
Moreover, if there is at most one resolver,
then all outputs are the same.

Formally, the protocol accepts values in some set $V$ as inputs and 
exports operations $\textit{propose}(v)$, $v\in V$, and 
$\textit{resolve}()$ that, once called by a process, indicates that
the process becomes a resolver for RAP. 
The propose operation returns some value in $V$, 
and the following guarantees are provided:    
(i) Every returned value is a proposed value; 
(ii) If all processes start with the same input value or some
  process returns, then 
  every correct process returns;
(iii) If a correct process becomes a resolver, then 
 every correct process returns;	 
(iv) If at most one process becomes a
  resolver, then at most one value is returned.

\begin{figure}[tbp]
\hrule \vspace{1mm} {\footnotesize
\begin{tabbing}
 bbb\=bb\=bb\=bb\=bb\=bb\=bb\=bb \=  \kill
Shared variables: \\
\> $D$, initially $\bot$\\
Local variables: \\
\> $\id{resolver}$, initially $\false$\\
\vspace{2mm}
\textit{propose}$(v)$\\ 
\nnll\>  $(\textit{flag},\id{est}) := \id{CA.propose}(v)$\\
\nnll\> {\bf if} $\textit{flag}=\id{commit}$ {\bf then}\\
\nnll \label{line:commit}\>\> $D:=\id{est}$; \textit{return}($\id{est}$)\\ 
\nnll\label{line:wait1}\> {\bf repeat }\\
\nnll\label{line:resolve}\>\>  {\bf if} \id{resolver} {\bf then} $D:=\id{est}$ \\  
\nnll\label{line:wait2}\> {\bf until} $D\neq\bot$\\
\nnll\> \textit{return}($D$)\\ 
\vspace{2mm}
\textit{resolve}$()$\\ 
\nnll\>  $\id{resolver} := \true$
\end{tabbing}
\vspace{-2mm} 
\hrule 
}
\caption{Resolver agreement protocol: code for each process}
\label{fig:rap}
\end{figure}

A protocol that solves RAP is presented in Figure~\ref{fig:rap}. 
The protocol uses the \emph{commit-adopt} abstraction (CA)~\cite{Gaf98} 
exporting one operation $\textit{propose}(v)$ that returns $(\id{commit},v')$
or $(\id{adopt},v')$, for $v,v'\in V$, 
and guarantees that (a) every returned value is a proposed value, (b) 
if only one value is proposed then this value must be
committed, 
(c) if a process commits a value $v$, then every process that returns
adopts $v$ or commits $v$, and (d) every correct process returns.        
The commit-adopt abstraction can be implemented wait-free
~\cite{Gaf98}.
 
In the protocol, a process that is not a resolver takes
a finite number of steps and then either returns with a value, or waits
on one posted in register $D$ by another process or by a resolver. A process that waits for
 an output (lines~\ref{line:wait1}-\ref{line:wait2})
considers the agreement protocol \emph{stuck}. 
An agreement protocol for which a value was posted in $D$ is called
\emph{resolved}. 

\begin{lemma}
\label{lem:rap}
The algorithm in Figure~\ref{fig:rap} implements RAP.
\end{lemma}
\begin{proof}
Properties (i) and (ii) follow from the properties of CA and the algorithm: every
returned value is a proposed value and if all inputs are $v$ or some
process returns (after writing a non-$\bot$ value $v$ in $D$), 
then every process commits on $v$ and returns $v$ in line~\ref{line:commit}.

If there is a correct resolver, it eventually writes some value in $D$
(line~\ref{line:resolve}), and eventually every other process returns
some value, and thus property (iii) holds. 

Moreover, a returned value
either was committed in an instance of CA or was written to $D$ by a resolver. 
Even if some process returned a value $v$ committed in CA, then by the
properties of CA, the only value that a resolver can write in $D$ is $v$. 
Thus, if there is at most one resolver, the protocol can return at most one value, 
and property (iv) holds.     
\end{proof}

%

\section{From wait-freedom to $\L$-resilience}
\label{sec:wfl}
%
Suppose that $T_{\L}$ is weakly wait-free solvable 
and let $A_{\L}$ be the corresponding wait-free protocol.
We show that weak wait-free solvability of $T_{\L}$ implies
weak $\L$-resilient solvability of $T$ by presenting an algorithm $A$ 
that uses $A_{\L}$ to solve $T$ in every $\L$-resilient execution.

First we describe the \emph{doorway} protocol (DW), the only $\L$-dependent 
part of our transformation. 
The responsibility of DW is
to collect at each process a subset of the inputs of $T$ 
so that all the collected subsets constitute 
a legitimate input vector for task $T_{\L}$ (property (1) in Section~\ref{sec:wft}).
The doorway protocol does not require the knowledge
of $T$ or $T_{\L}$ and depends only on $\L$.

In contrast, the second part of the transformation described in Section~\ref{sec:wflsim}
does not depend on $\L$ and is implemented by simply invoking the wait-free task 
$T_{\L}$ with the inputs provided by DW. 

\subsection{The doorway protocol}  
\label{sec:dw}
Formally, a DW protocol ensures that in every $\L$-resilient execution
with an input vector $I\in \I$,  
every correct participant eventually obtains a \emph{set} of inputs of
$T$ so that the resulting input vector $I'\in\T_{\L}$ 
complies with property (1) in Section~\ref{sec:wft} with respect to $I$.  
 \begin{figure}[tbp]
\hrule \vspace{1mm} {\footnotesize
\begin{tabbing}
 bbb\=bb\=bb\=bb\=bb\=bb\=bb\=bb \=  \kill
Shared variables: \\
\> $R_j$, $j=1,\ldots,n$, initially $\bot$\\
Local variables: \\
\> $S_j$, $j=1,\ldots,\HSS(\Pi,\L)$, initially $\emptyset$\\
\> $\ell_j$, $j=1,\ldots,\HSS(\Pi,\L)$, initially $0$\\
\vspace{2mm}
\nnll\>  $R_i := $ input value of $T$\\
\nnll\label{line:atomic}\>  {\bf wait until} $\id{snapshot}(R_1,\ldots,R_n)$ contains
inputs for some set in $\L$\\
\nnll\> {\bf while} \id{true} {\bf do}\\
\nnll\label{line:cycle1}\>\> $S:= \{p_i\in P, R_i\neq\bot\}$ \hspace{1cm} \{the current participating set\}\\ 
\nnll\>\> {\bf if} $p_i\in H^S$ {\bf then} \hspace{1cm} \{$H^S$ 
				is deterministically chosen in $\HS(S,\L)$\}\\
\nnll\>\>\> $m$ := the index of $p_i$ in $H^S$ \\
\nnll\>\>\> $\id{RAP}_m^{\ell_m}.\id{resolve}()$\\
\nnll\label{line:foras}\>\> {\bf for} each  $j=1,\ldots,|H^S|$ {\bf do}\\ 
\nnll\>\>\> {\bf if} ${\ell_j=0}$ {\bf then}\\
\nnll\label{line:hs}\>\>\>\>  $S_j:= S$ \\ 
\nnll\label{line:rap}\>\>\> take one more step of $\id{RAP}_j^{\ell_j}.\id{propose}(S_j)$\\
\nnll\>\>\> {\bf if} $\id{RAP}_j^{\ell_j}. \id{propose}(S_j)$ returns $v$ {\bf then}\\
\nnll\label{line:ca}\>\>\>\>  $(\id{flag},S_j):= \id{CA}_{j}^{\ell_j}.\id{propose}(v)$\\
\nnll\>\>\>\> {\bf if} $(\id{flag}=\id{commit})$ {\bf then}\\
\nnll\>\>\>\>\> \id{return}$(\{(s,R_{s})\}_{p_{s}\in S_j})$    
	\hspace{1cm} \{return the set of inputs of processes in $S_j$\}\\
 \nnll\label{line:cycle2}\>\>\>\> $\ell_j:=\ell_j+1$
\end{tabbing}
\vspace{-2mm} 
\hrule 
}
\caption{The doorway protocol: the code for each process $p_i$}
\label{fig:dw}
\end{figure}

The algorithm implementing DW is presented in Figure~\ref{fig:dw}.
Initially, each process $p_i$ waits until it collects inputs for a
set of participating processes that includes at least one live set.  
Note that different processes may observe different  participating sets.
Every participating set $S$ is  
associated with $H^S\in\HS(S,\L)$, some deterministically chosen
hitting set of $(S,\L)$.
We say that $H^S$ is a \emph{resolver set for $\Pi$}:
if $S$ is the participating set, then we initiate $|H^S|$ 
parallel sequences of agreement protocols with resolvers. 
Each sequence of agreement protocols can return at most one value
and we guarantee that, eventually, every sequence is associated 
with a distinct resolver in $H^S$.   
In every such sequence $j$, each process $p_i$ sequentially goes through 
an alternation of RAPs and CAs (see Section~\ref{sec:rap}): 
$\id{RAP}_j^1,\id{CA}_j^1,\id{RAP}_j^2,\id{CA}_j^2,\ldots$.
The first RAP is invoked with the initially observed set of participants, and
each next CA (resp., RAP) takes the output of the previous RAP (resp., CA)
as an input.
If some $\id{CA}_j^{\ell}$ returns $(\id{commit},v)$,
then $p_i$ returns $v$ as an output of the doorway protocol.


\begin{lemma}
\label{lem:dw}
In every $\L$-resilient execution of the algorithm in Figure~\ref{fig:dw} 
starting with an input vector $I$, every correct process $p_i$ terminates with 
an output value $I'[i]$, and the resulting vector $I'$ complies with 
property (1) in Section~\ref{sec:wft} with respect to $I$. 
\end{lemma}
\begin{proof}
Consider any $\L$-resilient execution of the algorithm.
We say that an agreement sequence $j$ is \emph{triggered} if 
some process $p_i$ accessed $\id{RAP}_j^1$, the first RAP instance in the
sequence, in line~\ref{line:rap}.
First, we observe that if a sequence $j$ is triggered, then value
$S_j$ proposed to its first RAP instance
by any process $p_i$
(we simply say $p_j$ \emph{proposes  $S_j$ to sequence $j$}) 
is such that $S_j$ is a set of participants containing a live set and $\HSS(S_j,\L)\geq j$
(lines~\ref{line:atomic} and ~\ref{line:cycle1}--\ref{line:hs}).
Recall that for all process subsets $S$ and $S'$ such that
$S\subseteq S'$, we have $\HSS(S,\L)\leq \HSS(S',\L)|$.
By the properties of atomic snapshot
(line~\ref{line:atomic}), for every two sets $S_j$ and $S_{\ell}$
proposed to sequences  $j$ and $\ell$ such that $j<\ell$, we have
$S_j\subset S_{\ell}$.

Consider any $\L$-resilient execution of the algorithm.
Let $S$ be the set of participants in that execution.
Every value returned by the protocol must be committed in
some $\id{CA}_{j}^{\ell_j}$, $1\leq j \leq |H^S|$ (line~\ref{line:ca}).
By the properties of CA, every committed value 
is adopted by every process and then proposed to the 
next instance of RAP (line~\ref{line:rap}). 
By the properties of CA and RAP, every value returned by an instance of 
CA or RAP was previously proposed to the instance, and thus, 
no two different values can be returned 
in a given agreement sequence.
Let $m$ be the highest agreement sequence in which some value $\bar S_m$ was returned. 
Thus, at most $m$ distinct sets $\bar S_1$, $\dots$, $\bar S_m$ are returned in total and all of
these sets are subsets of $\bar S_m$.
Thus, $\cup_{j,I'[j]\neq\bot} \bar S_j = \bar S_m$.
Recall that $\HSS(\cup_{j,I'[j]\neq\bot} \bar
S_j,\L)=\HSS(\bar S_m,\L)\geq m$ 
and property (1b) holds. 
Finally, resulting $I'$ is an image of $I$ with respect to
some sequence $S_1,\ldots, S_m$ where each $S_j$ is a superset of 
a live set, and property (1a) also holds.  
 
To show liveness, we first observe that 
in an $\L$-resilient execution, line~\ref{line:atomic} is non-blocking.
Further, the body of the cycle in lines~\ref{line:cycle1}--\ref{line:cycle2}
contains no blocking statements.
Thus, every correct process returns or goes through infinite number of 
cycles, trying to 
advance all triggered agreement sequences $1,\ldots,|H^S|$,
where $S$ is the  participating set.

To prove that every correct process terminates, 
it is sufficient to show that at least 
one process returns. 
Indeed, suppose that a process $p_i$ returns after having committed on a set $S_j$ in 
some $\id{CA}_j^{\ell_j}$ (line~\ref{line:ca}).
If a process returns from an instance of RAP, 
then every correct process returns from the instance (property (ii) of RAP).
Also, every correct process returns from each instance of CA.
Thus, every correct process eventually reaches $\id{CA}_j^{\ell_j}$.
By the properties of CA, every process that returns in 
$\id{CA}_j^{\ell_j}$, adopts or commits $S_j$.
By properties (i) and (ii) of RAP, every correct process returns $S_j$
in $\id{RAP}_j^{\ell_j+1}$.
By the properties of CA, every correct process commits $S_j$ 
in $\id{CA}_j^{\ell_j+1}$, and returns.

Suppose, by contradiction that no process ever returns.
Eventually, all correct processes find the same set of participants
$S$ in line~\ref{line:cycle1} and, thus, agree on
the assigned hitting set $H^S$ of $(S,\L)$. 
In an $\L$-resilient execution, at most $|H^S|-1$ processes in $H^S$ can fail.
Otherwise, $H^S$ is not a hitting set, since it does not meet every
live set subset of $S$.
In a given agreement sequence $j$,  every $\id{RAP}_j^{\ell_j}$ is eventually associated 
with a distinct resolver in $H^S$.
Thus, by property (iii) of RAPs there exists an agreement sequence $j\in\{1,\ldots,|H^S|\}$, 
that is eventually associated with a distinct correct resolver $p_r$ in $H^S$.
Since, eventually, $p_r$ is the only resolver of RAPs in sequence $j$
and, by our assumption, agreement sequence $j$ goes through an infinite number of RAP instances,
there is an instance $\id{RAP}_j^{\ell_j}$ in which $p_r$ is the only resolver and,
by property (iv) of RAPs, exactly one value $S_j$ is returned to every correct process.
Thus, every correct process commits on $S_j$ in $\id{CA}_j^{\ell_j}$ and returns 
	--- a contradiction.
\end{proof}

\subsection{Solving $T$ through the doorway}
\label{sec:wflsim}
Given the DW protocol described above, it is straightforward
to solve $T$ by simply invoking $A_{\L}$ with the inputs provided by DW. 
Thus:
\begin{theorem}
\label{th:wfl}
Task $T$ is weakly $\L$-resiliently solvable if $T_{\L}$ is
weakly wait-free solvable. 
\end{theorem}
\begin{proof}
By Lemma~\ref{lem:dw}, every execution of DW starting with an input vector $I$ 
makes sure that each process is assigned a set of inputs of $T$ for
some participating live set, and property (1) of $T_{\L}$ is satisfied with respect to $I$
and the resulting vector $I'$.   
Now we use $A_{\L}$ with $I'$, and, 
by the property (2) of $T_{\L}$, at least one participating set in $\L$
obtains outputs.
\end{proof}

%
%

\section{From $\L$-resilience to wait-freedom}
\label{sec:lwf}
%
Suppose $T$ is weakly $\L$-resiliently solvable, and let $A$ be the
corresponding protocol. 
We describe a protocol $A_{\L}$ that
solves $T_{\L}$ by wait-free simulating an $\L$-resilient 
execution of $A$. 

For pedagogical reasons, we first present  
a simple \emph{abstract simulation} (AS)
technique. AS captures the intuition that 
a group of 
simulators sharing the initial view of the set of 
participating simulated codes
should appear as a single simulator. 
Therefore, 
an arbitrary number of simulators starting with $j$ distinct
initial views
should be able to simulate a $(j-1)$-resilient execution.

Then we describe our specific simulation and show that it is an instance
of AS, and thus it indeed generates a $(j-1)$-resilient execution of
$\L$, where $j$ is the number of distinct inputs of $T_{\L}$.   
By the properties of $T_{\L}$, we immediately obtain a desired $\L$-resilient
execution of $A$. 

\remove{ Petr:move to Sec 5.2?
Indeed, suppose that the result of every read operation of the code of $A$ is agreed upon among 
the simulators using a RAP instance.
We say that a code of $a$ is \emph{triggered} if some participating simulator 

Simulators proceed in simulating the codes of $A$ in a round-robin fashion: 
first read command of the first triggered code 

If all simulators start with the same set of inputs to $A$ they move
Round-Robin from $read$ command to the next, each read resolved by $RAP$. Since they all move in the same 
order all propose the same value for a $read$, and the simulation will progress with all simulated codes with
inputs progressing Round-Robin. What is another group of simulators joins in with another set of inputs. We would like
to go Round-Robin over the larger group, but until coordination is achieved simulators may propose different
values to a $read$. The idea behind the Abstract Simulation is to show that in this case only one $read$ may
be ``problematic'' and in need of a resolver. If and when resolved, it may cause another ``problematic'' $read$.
The conclusion since there may be only single problematic $read$ that need a resolver, at most one simulated 
thread of $A$ with an input will not progress.

Of course we use two groups as an example. In the case of $j$ groups, at most $j-1$ $read$s may become problematic
preventing the progress of at most $j-1$ threads of $A$.
}
\subsection{Abstract simulation}
\label{sec:abstract}
%
Suppose that we want to simulate a given $n$-process protocol,
with the set of codes $\{\code_1,\ldots,\code_n\}$.    
Every instruction of the simulated codes (read or write) is associated with a
unique \emph{position} in $\mathbb{N}$.
E.g., we can enumerate the instructions as follows: the first instructions of each simulated code, 
then the second instructions of each simulated code, etc.\footnote{In
  fact, only read instructions of a read-write
  	protocol need to be simulated since these are the only steps that
	may trigger more than one state transition of the invoking process~\cite{BG93b,BGLR01}.} 

A \emph{state} of the simulation is a map of the set of positions to \emph{colors} $\{U,IP,V\}$,
every position can have one of three colors: $U$ (\emph{unvisited}), $IP$ (\emph{in progress}),
or $V$ (\emph{visited}).  
Initially, every position is unvisited.
The simulators share a function \id{next} that maps every state 
to the next unvisited position to simulate.
\emph{Accessing} an unvisited position by a simulator 
results in changing its color to $IP$ or $V$.  

\begin{figure}[tbph]
  \centering
  \includegraphics[scale=0.4]{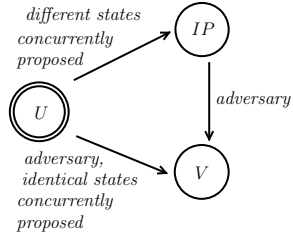}
  \caption{State transitions of a position in AS.}
  \label{fig:states}
\end{figure}

The state transitions of a position are summarized
in Figure~\ref{fig:states}, and the rules the simulation follows are
described below:
%
{\footnotesize
\begin{enumerate}
\item[(AS1)] Each process takes an atomic snapshot of the current state 
$s$ and goes to position $\id{next}(s)$ \emph{proposing} state $s$.
For each state $s$, the color of $\id{next}(s)$ in state $s$ is $U$.\\ 
- If an unvisited position is \emph{concurrently} 
accessed by two processes proposing different states, 
then it is assigned color $IP$.\\
- If an unvisited position is accessed 
by  every process proposing the same state, 
it may only change its color to $V$.\\
- If the accessed position is already $V$ (a faster process 
accessed it before), then the process leaves the position unchanged, 
takes a new snapshot, and proceeds to the next position.

\item[(AS2)] At any point in the simulation, the \emph{adversary} may take an in-progress ($IP$) 
position and atomically 
turn it into $V$ or take a \emph{set} of unvisited ($U$) 
positions and atomically turn them into $V$.

\item[(AS3)] Initially, every position is assigned color $U$.
The simulation starts when the adversary changes 
colors of some positions to $V$. 
\end{enumerate}
}
%
We measure the progress of the simulation by the number of positions
turning from $U$ to $V$. Note that by changing $U$ or $IP$ positions to $V$, 
the adversary can potentially hamper the simulation, by causing 
some $U$ positions to be accessed with different states and thus 
changing their colors to $IP$.
However, the following invariant is preserved:
\begin{lemma}
\label{lem:abstract}
If the adversary is allowed 
at any state to change the colors of arbitrarily many $IP$ positions to $V$,
and throughout the simulation has $j$ chances 
to atomically change any set of $U$ positions to $V$,
then at any time 
there are at most $j-1$ $IP$ positions.
\end{lemma}
\begin{proof}
Note that 
in the periods when the adversary does not move, 
every new accessed position may only become visited. 
Indeed, even though the processes run asynchronously, 
they march through the same sequence of snapshots.
Every snapshot a process takes is either a fresh view that points to 
a currently unvisited position, or was previously observed 
by some process and it points to a visited position. 
In both cases, no new $IP$ position can show up.

Now suppose that the adversary changed the  color of a position from $IP$ to $V$, 
thus decreasing the number of $IP$ positions by one.
This may result in one distinct inconsistent (not seen by any
other simulator) state that points (through function \id{next})
to one distinct position. 
Thus, at most one position can be accessed with diverging states, resulting 
in at most one new $IP$ position. Thus, in the worst case, the total number of $IP$
positions remains the same.  

Now suppose that $j$ sets of positions changed their colors from $U$ to $V$, 
one set at a time.
The change of colors of the very first group starts the simulation and thus 
does not introduce $IP$ positions.
Again, every subsequent group of changes can result in at most one inconsistent state, 
which may bring up to $j-1$ new $IP$ positions in total.
\end{proof}

\subsection{Solving $T_{\L}$ through AS}   
\label{sec:lwfsim}
Now we show how to solve $T_{\L}$ by simulating a protocol $A$ that weakly $\L$-resiliently 
solves $T$. First, we describe our simulation and show that it
instantiates AS, which allows 
us to apply Lemma~\ref{lem:abstract}.

Every simulator $s_i\in\{s_1,\ldots,s_n\}$ posts its input in the shared memory and then 
continuously simulates participating codes in $\{\code_1,\ldots,\code_n\}$ of algorithm $A$ 
in the breadth-first manner:
the first command of every participating code, the second command of
every participating code, etc. 
(A code is considered participating if its input value has been posted by at least one simulator.)
The procedure is similar to BG-simulation, except that the result of every \id{read} command in the code is agreed upon
through a distinct RAP instance. 
Simulator $s_i$ is statically assigned to be the only resolver 
of every \id{read} command in $\code_i$.

The simulated read commands (and associated RAPs) are 
treated as positions of AS.
Initially, all positions are $U$ (unvisited).
The outcome of accessing a RAP instance of a position determines its color. 
If the RAP is resolved (a value was posted in $D$ in
line~\ref{line:commit} or~\ref{line:resolve}), then it is given color $V$ (visited).  
If the RAP is found stuck (waiting for
 an output in lines~\ref{line:wait1}-\ref{line:wait2}) 
by some process, then it is given color $IP$ (in progress). 
Note that no RAP accessed with identical proposals 
can get stuck (property (ii) in Section~\ref{sec:rap}).
After accessing a position, the simulator chooses the first not-yet executed 
command of the next participating 
code in the round-robin manner (function \id{next}).
For the next simulated command, the simulator proposes its current view of the simulated state,
i.e., the snapshot of the results of all commands simulated so far (AS1).

Further, if a RAP of $\code_i$ is observed stuck by a simulator 
(and thus is assigned color $IP$), but later gets resolved by $s_i$, 
we model it as the adversary spontaneously  changing the position's color 
from $IP$ to $V$.
Finally, by the properties of RAP, a position can get color $IP$ 
only if it is concurrently accessed with diverging states (AS2).

We also have $n$ positions corresponding to the input values of the codes,
initially unvisited.
If an input for a simulated process $p_i$ is posted by a simulator, the initial position of $\code_i$ turns into $V$.
This is modeled as the intrusion of the adversary, and 
if simulators start with $j$ distinct inputs, then the adversary 
is given $j$ chances to atomically change sets of $U$ positions to $V$.
The simulation starts when the first set of simulators post their inputs 
concurrently take identical snapshots (AS3).


Therefore, our simulation is an instance of AS, 
and thus we can apply Lemma~\ref{lem:abstract} to prove the following result:
\begin{lemma}
\label{lem:inv}
If the number of distinct values in the input vector of $T_{\L}$ is $j$, then 
the simulation above blocks at most $j-1$ simulated codes.
\end{lemma} 
\begin{proof}
Every distinct value $S$ in an input vector of $T_{\L}$ posted by a participating simulator 
results in the adversary changing some set of initial positions from $U$ to $V$.
(Note that the set can be empty if the inputs for set $S$ has been previously
posted by another simulator.)
By Lemma~\ref{lem:abstract}, at any time there are at most $j-1$ $IP$ positions, 
i.e., at most $j-1$ RAPs for read steps that are stuck.
Thus, in the worst case, at most $j-1$ simulated codes can block forever.
\end{proof}
The simulated execution terminates when some simulator observes
outputs of $T$ for at least one participating live set.  
Finally, using the properties of the inputs to task $T_{\L}$
(Section~\ref{sec:wft}), we derive that eventually, some 
participating live set of simulated processes obtain outputs.
Thus, using Theorem~\ref{th:wfl}, we obtain:
\begin{theorem}
\label{th:lwf}
$T$ is weakly $\L$-resiliently solvable if and only if $T_{\L}$ is weakly wait-free solvable.
\end{theorem}
\begin{proof}
Suppose that we are given an input vector $I'$ of $T_{\L}$ with $j$ distinct
values, each value consists of inputs of $T$ for a set of processes containing a live set. 
By property (1a) of $T_{\L}$ (Section~\ref{sec:wft}), these input sets are consistent with some 
input vector $I$ of $T$.
We call the set of simulated processes that obtain inputs of $T$ the
\emph{participating set} of $T$, and denote by $\Pi'$.

Since every simulated step goes through a RAP with a single resolver,
by property (iv)  of RAP (Section~\ref{sec:rap}), simulators agree on the result 
of every simulated read command,
and thus we simulate a correct execution of algorithm $A$ (solving $T$).   
    
By Lemma~\ref{lem:inv}, at most $j-1$  processes can fail in the
simulated execution of $A$.
By property (1b) of $T_{\L}$, the size of the hitting set of the participating set
$H(\Pi',\L)$ is at least $j$. 
Thus, there is at least one live set in $\Pi'$ that contains no faulty
simulated process. 
This live set accepts infinitely many steps in the simulated
execution of $A$ and, by weak $\L$-resilient solvability, must
eventually output. 
This set of outputs constitutes the output of $T_{\L}$.
Since the output comes from an execution of $A$ 
starting with $I$, the output satisfies property (2) of $T_{\L}$.

Thus, the algorithm indeed solves $T_{\L}$.
\end{proof}



\section{Related work}
\label{sec:related}

\ignore{ 
This work touches two topics: relating $t$-resilience and wait-freedom and 
characterizing conditions under which 
distributed tasks can be solved.    
}
%
The equivalence between $t$-resilient task solvability and wait-free
task solvability has been initially established for colorless tasks
in~\cite{BG93b,BGLR01},
and then extended to all tasks in~\cite{Gaf09-EBG}.
\remove{
In short, the extended BG-simulation~\cite{Gaf09-EBG} goes through
establishing $T_{\L}$, a ``variant'' of a task $T$ defined on $t+1$ simulators,
where each simulator is given values for at least $n-t$ processes in $T$ so that 
when all $t+1$ simulators participate, every process in $T$ is given an input.
Respectively, every simulator is supposed to produce outputs for at least 
$n-t$ processes in $T$ so that when $j$ simulators terminate ($j=1,\ldots,t+1$), 
at least $n-t-1+j$ processes in $T$ obtain outputs.
}
In this paper, we consider a wider class of assumptions than
simply $t$-resilience, which can be seen as a strict generalization of~\cite{Gaf09-EBG}. 
\remove{
For general $\L$, we restrict ourselves to weak termination, but in the $t$-resilient case, 
as we sketch in the next section,
our results imply an alternative derivation of the characterization in~\cite{Gaf09-EBG}.
}

%
Generalizing $t$-resilience, 
Janqueira and Marzullo~\cite{JM07-cores} considered the case of \emph{dependent} 
failures and proposed describing the allowed executions through \emph{cores} and 
\emph{survivor sets} which roughly translate to our hitting sets 
and live sets.
Note that the set of survivor sets (or, equivalently, cores) 
exhaustively describe only superset-closed adversaries.
\ignore{
A core, a minimum size set of processes that
cannot be all faulty in any execution, roughly translates to a hitting
set of an $\L$-resilient adversary. A survivor set $S$ is a minimal set of
processes that contains only correct processes.
In contrast, maintaining a live set $L\in\L$ means that we
consider all executions in which $L$ is a \emph{subset} of the correct
set.  
}
More general adversaries introduced by Delporte et al.~\cite{delporte09adversary}
	are defined as a set of exact correct sets.
It is shown in~\cite{delporte09adversary} that 
	the power of an adversary $\A$ to solve colorless tasks 
	is characterized by $\A$'s \emph{disagreement power}, 
	the highest $k$ such that $k$-set agreement cannot be solved assuming $\A$:
	a colorless task $T$ is solvable with 
	adversary $\A$ of disagreement power $k$ 
	if and only if it is solvable $k$-resiliently.
Herlihy and Rajsbaum~\cite{HR10} (concurrently and
        independently of this paper) derived this result 
	for a restricted set of superset-closed 
	adversaries with a given core size using elements of modern combinatorial 
        topology.
Theorem~\ref{th:colorless} in this paper derives this result directly,
	using very simple algorithmic arguments.   

Considering only colorless tasks is a strong restriction, 
since such tasks allow for definitions that only depend on 
\emph{sets} of inputs and \emph{sets} of outputs, regardless of which
processes actually participate.
(Recall that for colorless tasks, solvability and our weak solvability are
equivalent.)
The results of this paper hold for all tasks. On the other hand,  as~\cite{HR10},
we only consider the class of superset-closed adversaries.
This filters out some popular liveness properties,
such as \emph{obstruction-freedom}~\cite{HLM03}.
Thus, our contributions complement but do not contain the results 
in~\cite{delporte09adversary}.
A protocol similar to our RAP was earlier proposed in~\cite{IR09-rap}.

\remove{
Our direct reduction of $\L$-resilience to wait-freedom seems 
to be the only option for tasks whose correctness conditions depend on the participating 
set. Indeed, for such tasks $k$-resilient solvability may not tell enough to 
characterize the task completely, and, intuitively, this is the reason why 
indirect characterizations of~\cite{delporte09adversary} do not apply.
}

\section{Side remarks and open questions}
\label{sec:discussion}
%
\textbf{Doorways and iterated phases.} Our characterization shows 
an interesting property of weak $\L$-resilient solvability:
    To solve a task $T$ weakly $\L$-resiliently, we can
    proceed in two logically synchronous phases. In the
    first phase, processes wait to collect ``enough'' input values,
    as prescribed by $\L$, without knowing anything about $T$.
    Logically, they all finish the waiting phase simultaneously.
    In the second phase, they all proceed wait-free to produce a solution.
    As a result, no process is waiting on another process that already proceeded to the
    wait-free phase. Such phases are usually referred to as iterated phases \cite{BG97}.
    In~\cite{Gaf09-EBG},
    some processes are waiting on others to produce an output
    and consequently the characterization in~\cite{Gaf09-EBG}
    does not have the iterated structure.

\noindent
\textbf{$\L$-resilience and general adversaries.} 
The power of a general adversary of~\cite{delporte09adversary} 
is not exhaustively captured by its
hitting set. 
In a companion paper~\cite{GK10-OPODIS}, we propose a simple characterization of 
the set consensus power of 
a general adversary $\A$ based on the hitting set sizes of its recursively 
proper subsets. 
Extending our equivalence result to general adversaries and 
getting rid of the weak solvability assumption
are two challenging open questions.

{\small 
\bibliography{references}
}

\end{document}